\documentclass[pdflatex,sn-mathphys-num]{sn-jnl}

\usepackage{graphicx}%
\usepackage{multirow}%
\usepackage{amsmath,amssymb,amsfonts}%
\usepackage{amsthm}%
\usepackage{mathrsfs}%
\usepackage[title]{appendix}%
\usepackage{xcolor}%
\usepackage{textcomp}%
\usepackage{manyfoot}%
\usepackage{booktabs}%
\usepackage{algorithm}%
\usepackage{algorithmicx}%
\usepackage{algpseudocode}%
\usepackage{listings}%
\usepackage{caption}
\usepackage{soul}

\theoremstyle{thmstyleone}%

\theoremstyle{thmstyletwo}%

\theoremstyle{thmstylethree}%

\begin{document}

\title[Time-Resolved Stress Analysis of Tissue Simulants During Needle-Free Jet Injection]{Time-Resolved Stress Analysis of Tissue Simulants During Needle-Free Jet Injection}

\author[1]{\fnm{Kohei} \sur{Yamagata}}
\author[1]{\fnm{Prasad} \sur{Sonar}}
\author[2]{\fnm{Kazuhiro} \sur{Terai}}
\author[1,3]{\fnm{Yuto} \sur{Yokoyama}}
\author*[1]{\fnm{Yoshiyuki} \sur{Tagawa}}\email{tagawayo@cc.tuat.ac.jp}

\affil[1]{\orgdiv{Department of Mechanical Systems Engineering}, \orgname{Tokyo University of Agriculture and Technology}, \orgaddress{\street{2-24-16, Naka-cho}, \city{Koganei}, \postcode{184-8588}, \state{Tokyo}, \country{Japan}}}

\affil[2]{\orgdiv{Research Center, R\&D Headquarters}, \orgname{Daicel Corporation}, \orgaddress{\street{JR Shinagawa East Bldg., 2-18-1, Konan, Minato-ku}, \city{Tokyo}, \postcode{108-8230}, \country{Japan}}}

\affil[3]{\orgdiv{Present address: Splash Lab, Mechanical Engineering Program}, \orgname{King Abdullah University of Science and Technology (KAUST)}, \orgaddress{\city{Thuwal}, \postcode{23955-6900}, \country{Saudi Arabia}}}

\abstract{Needle-free jet injection generates transient internal stress fields that can influence tissue deformation, pain-related stimulation, and cellular-level mechanical responses. However, the penetration mechanics have often been inferred from cavity deformation and interpreted mainly as shear-dominated behavior. In this study, high-speed photoelastic measurements were used to visualize and quantify optically integrated stress responses in a 5~wt\% gelatin tissue simulant during penetration by two needle-free injectors with different actuation mechanisms: the Actranza Lab, a pyro-drive injector driven by cartridge-based combustion, and the Biojector 2000, a commercially available CO$_2$-driven injector. A polarization camera operated at 60{,}000~fps was used to obtain the phase difference and principal stress orientation, allowing evaluation of the photoelastic stress-intensity response and its decomposed normal- and shear-stress-related components. Under the same injection volume of 20~$\mu$L, the Actranza Lab formed a narrow, depth-oriented cavity, whereas the Biojector 2000 produced a wider, bulged cavity. In both cases, a clear normal-stress-difference component developed around the cavity. This component became comparable to or greater than the shear-stress component for the Actranza Lab and became dominant during the later cavity-bulging stage for the Biojector 2000. These results show that needle-free jet penetration cannot be described solely by shear stress; instead, injector-dependent cavity dynamics generate multi-component tissue loading. The findings provide an engineering basis for evaluating needle-free injector performance and for designing systems that improve delivery while reducing mechanical burden on tissue.

}

\maketitle

\section{Introduction}
\label{sec:introduction}

Needle-free injectors are increasingly regarded as an alternative to conventional hypodermic needles because they may reduce the risk of needle-stick injuries and needle phobia~\cite{kane1999,nir2003,gill2007,mitragotri2006,arora2007,menezes2009,han2010,cu2020,baxter2005,romgens2016,schoppink2022}. In particular, liquid-jet-based needle-free injection systems have attracted growing interest, and a variety of jet-generation strategies have been developed. The mechanical interaction between a fluid jet and biological tissue is a key factor governing the effectiveness and safety of drug delivery. Parameters such as penetration speed, jet delivery capacity, and tissue loading are closely associated with drug dispersion, tissue damage, and pain perception~\cite{sweetman2011, meng2014,kedarasetti2020,rodriguez2018,van2015,bukavc2019fluid,sass2020,pahlavian2015,bluestein2017,doost2016,lawal2024}. In needle-free injectors, the jet-generation mechanism---such as compressed gas, spring actuation, pyrotechnic actuation, or cavitation-induced pressure waves---determines the pressure rise, delivery duration, flow-rate profile, maximum jet velocity, jet diameter, and jet stability. These factors strongly affect penetration depth, cavity morphology, and the resulting mechanical response of the target tissue. Previous studies have shown that controlling the jet-pressure profile enables control of penetration depth and injection volume in soft materials and skin tissue~\cite{miyazaki2019,sonoda2023promising,sonoda2025intradermal,inoue2023induction,shimizu2025new,taberner2012needle,stachowiak2009dynamic,moradiafrapoli2017high,rane2020computational,tagawa2012highly,rohilla2023focused,oyarte2020microfluidics}.

The penetration cavity is not merely a geometric consequence of jet impact, but a key feature governing liquid-jet penetration because its temporal evolution influences both penetration behavior and the surrounding mechanical response~\cite{miyazaki2021dynamic,shergold2006,van2023microfluidic}. In previous studies, the mechanics of liquid-jet penetration into soft materials have often been inferred from the observed cavity morphology and its temporal evolution. For example, liquid-jet penetration has been described in terms of cavity or crack formation in soft solids, while focused microjets have been shown to form narrow and elongated penetration paths that promote deep penetration~\cite{shergold2006,tagawa2013needle}. These studies have provided the basis for interpreting jet penetration as a process governed largely by localized deformation around the penetration path. In particular, shear-dominated deformation near the cavity wall has commonly been used to interpret the mechanical response associated with jet penetration. Miyazaki et al.~\cite{miyazaki2021dynamic} showed that a focused microjet forms a funnel-like cavity with a narrow penetrating tip and produces a relatively low-intensity shear-stress field that propagates as a shear-stress wave. In contrast, a non-focused microjet was associated with rapid cavity expansion and elastic rebound, producing a stronger compressive-stress response. These findings indicate that cavity dynamics are closely linked to the internal stress state and that the dominant stress mode can change depending on the penetration process. Nevertheless, the transient stress components generated around the cavity, especially the relative contributions of shear stress and the normal-stress difference, have not been fully quantified for practical needle-free injectors with different actuation mechanisms.

Recent studies have also indicated that needle-free jet injection can generate tissue-level shear and normal stresses associated with cavity formation, as well as cell-level shear stress induced by local flow within the tissue~\cite{sonoda2023promising}. In addition, because needle-free injection is often performed by pressing the nozzle against the skin prior to jet discharge, the tissue may be pre-compressed by the applied load. Under excessive loading conditions, such compression can inhibit liquid dispersion within the tissue~\cite{rohilla2020loading}. These mechanical stimuli may alter the local mechanical environment around cells and thereby contribute to intracellular molecular uptake and inflammatory or immune responses~\cite{yegutkin2000effect,chiu1997reactive,qin2015low}. For instance, a randomized clinical trial using the Biojector\textsuperscript{\textregistered} 2000\texttrademark{} demonstrated that, compared with conventional needle-and-syringe administration, needle-free delivery enhanced IFN-$\gamma$ ELISpot responses, CD8+ T-cell responses, and antibody responses after boosting following DNA vaccine priming, suggesting that the delivery modality can affect immunogenicity~\cite{graham2013dna}. Therefore, experimentally characterizing the relationship between jet behavior, cavity dynamics, and tissue loading is important for improving the efficacy and safety of needle-free injection devices.

Photoelastic measurement offers a useful approach for this purpose because it enables direct visualization and quantification of stress distributions within transparent soft materials. Photoelasticity is based on stress-induced birefringence, in which the refractive index of a material becomes direction-dependent owing to optical anisotropy under applied stress~\cite{yokoyama2026scaling,ramesh2016digital,yokoyama2024high}. The resulting refractive-index difference along two orthogonal directions produces a phase difference, which appears as a change in transmitted light intensity and enables visualization of the internal stress state. Conventional approaches for evaluating stress and strain in soft materials generally rely on marker introduction, such as fluorescent particles or characteristic surface patterns, followed by displacement tracking and estimation of interfacial stress and strain fields through mechanical modeling or finite-element analysis~\cite{li2022imaging,franck2007three,bergert2016confocal,hall2012mapping}. In contrast, photoelastic measurement directly captures the optical response associated with internal stress, allowing photoelastic stress-response distributions within the material to be visualized and quantified with high spatial resolution. Beyond penetration itself, the spatiotemporal distribution of tissue loading may influence drug accommodation, tissue damage, pain-related mechanical stimulation, and delivery performance. However, the transient internal stress topology generated by different needle-free injectors remains insufficiently characterized. Therefore, direct time-resolved measurement of the photoelastic stress response is important for linking injector mechanics to biomedical performance. In this study, photoelastic analysis is applied to compare two needle-free injectors driven by different actuation mechanisms by examining the photoelastic stress-response fields generated during liquid-jet penetration into skin-simulating materials. Particular attention is given to the relationship between injector-dependent cavity dynamics and the associated shear- and normal-stress-related responses. By linking penetration behavior to cavity dynamics and projected stress-response topology, this study provides mechanistic insights that may contribute to the rational design and improvement of needle-free injection devices.

The experimental procedure and measurement methodology are described in \S \ref{sec:expt}. The experimental results and discussion are presented in \S \ref{sec:results_discussion}. Finally, the conclusions are presented in \S \ref{sec:conclusion}.

\section{Experiments}
\label{sec:expt}

\subsection{Method}
\begin{figure}[t]
\centering
\includegraphics[width=\linewidth]{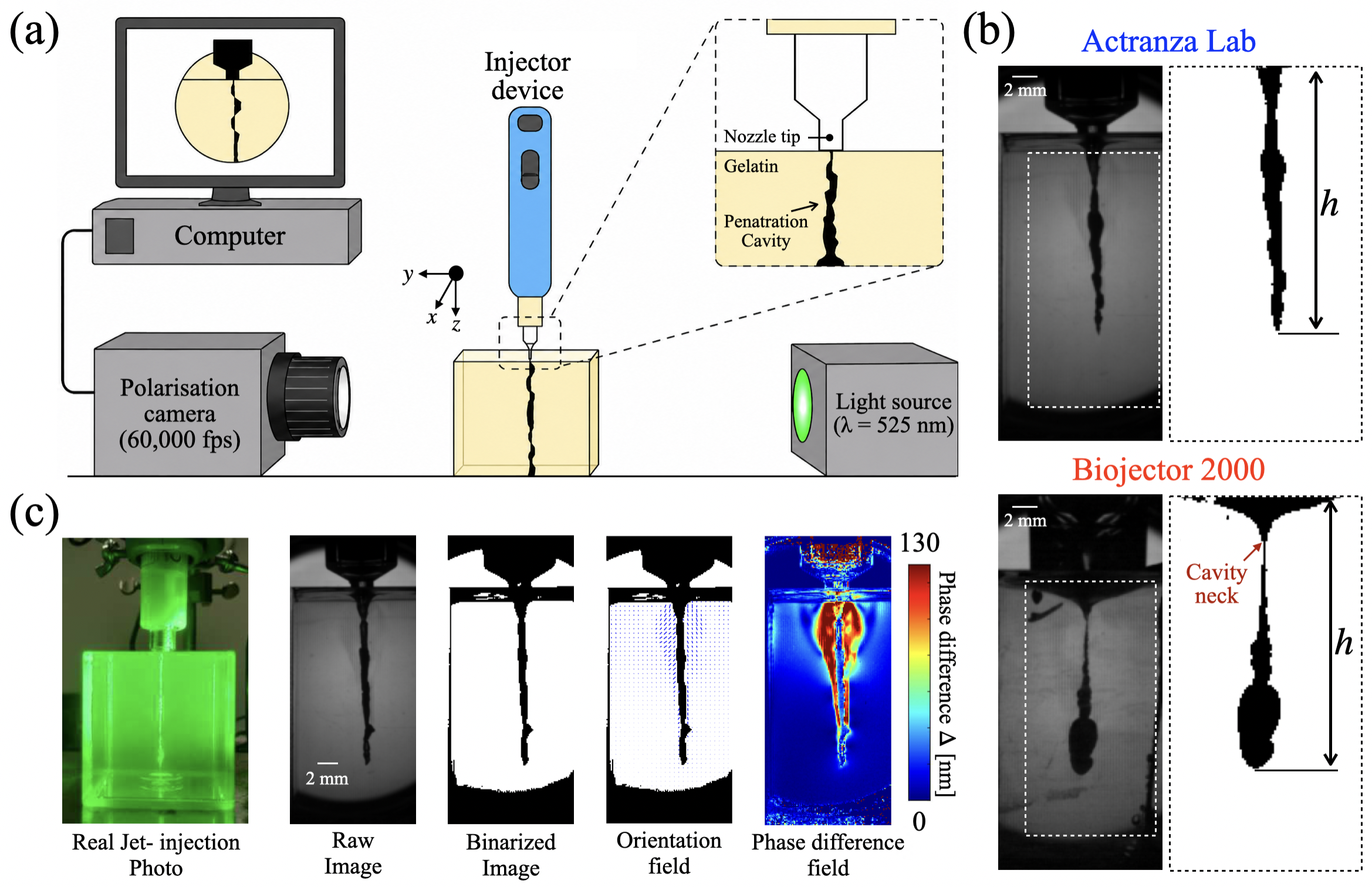}
\caption{
(a) Schematic of the experimental setup for high-speed photoelastic measurement during liquid-jet injection into gelatin.
(b) Representative raw images and corresponding binarized cavity images for the Actranza Lab and Biojector 2000 injectors.
The white dashed rectangles in the raw images indicate the $180\times332$ pixel region of interest (ROI) used for subsequent stress and cavity-geometry analyses.
The cavity depth $h$ was defined as the distance from the gelatin surface to the deepest point of the penetration cavity.
The cavity width $w$ was evaluated as the average horizontal extent of the binarized cavity within the ROI.
For the Biojector 2000, $w$ was evaluated only below the cavity neck indicated in the figure, excluding the neck region near the gelatin surface.
(c) Image-processing procedure from the recorded jet-injection image of the Actranza Lab to the raw image, binarized cavity image, orientation field, and phase-difference field.
The maximum magnitude of the phase difference is 130~nm.
}
\label{fig:schematic}
\end{figure}

\begin{table}[t]
\centering
\caption{Experimental specifications}
\label{tab:devices}
\begin{tabular}{@{}lll@{}}
\toprule
Experimental parameters and devices & Actranza Lab & Biojector 2000 \\
\midrule
Driving mechanism & Pyro-combustion & CO$_2$ cartridge \\
Injection volume, $V$ & 20~$\mu$L & 20~$\mu$L \\
Target material & Gelatin (5~wt\%) & Gelatin (5~wt\%) \\
Shear modulus, $G$ (Pa) & $ 5542$ & $ 5542$ \\
Nozzle diameter, $d_n$ (mm) & 0.1 & 0.17 \\
Nozzle length, $L_n$ (mm) & $\approx 1.0$ & $\approx 0.7$ \\
Delivery duration, $t_d$ (ms) & $\approx 10$ & $\approx 27$ \\
Average jet velocity, $U_{\mathrm{avg}}$ (m/s) & $\approx 250$ & $\approx 33$ \\
Average flow rate, $Q$ (mL/s) & $\approx 2.0$ & $\approx 0.74$ \\
\botrule
\end{tabular}
\end{table}

\begin{figure}[t]
\centering
\includegraphics[width=0.65\textwidth]{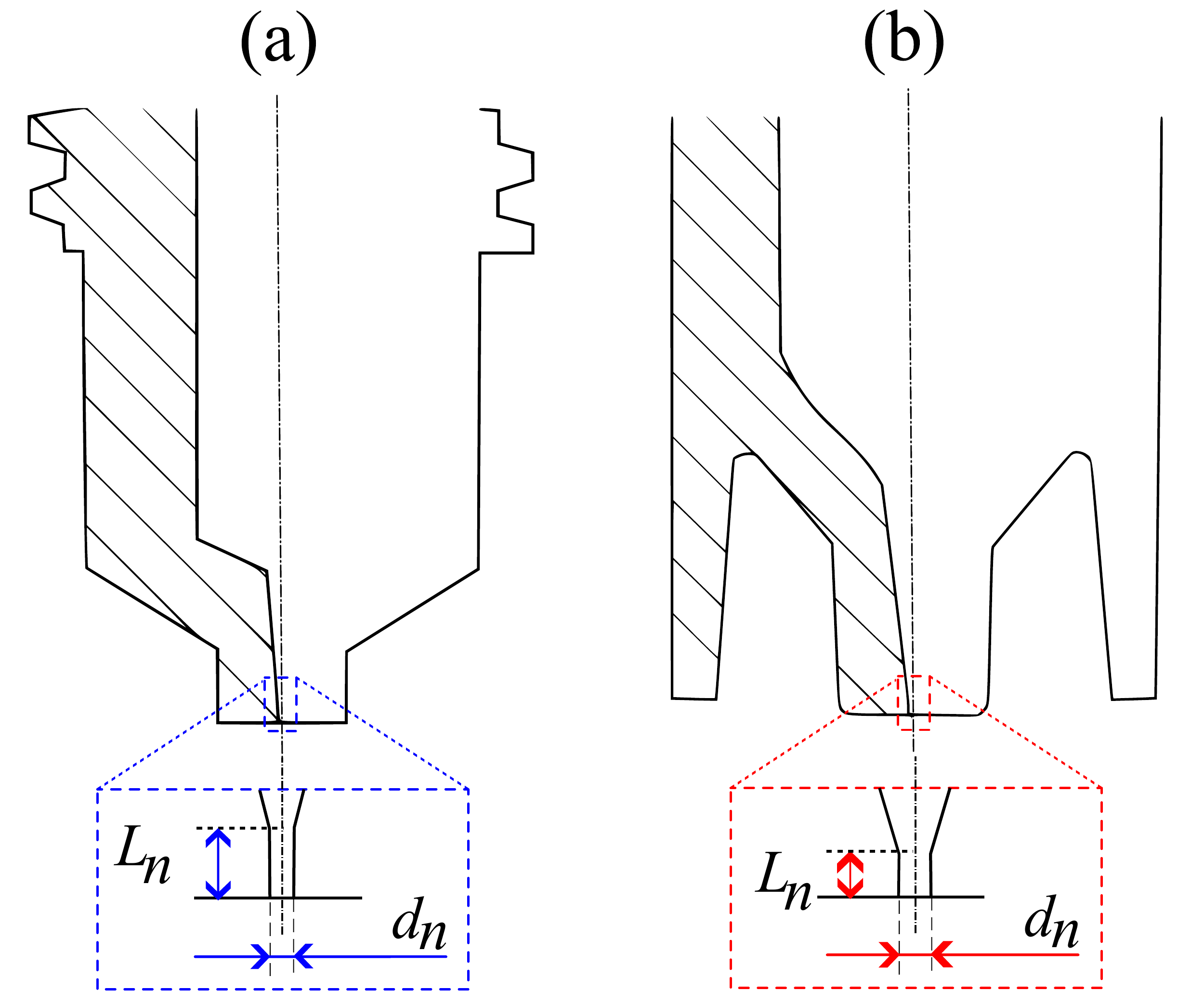}
\caption{Schematic of the nozzles of the two needle-free injectors used in this study: (a) Actranza Lab and (b) Biojector 2000. The insets indicate the nozzle parameters $L_n$ and $d_n$ listed in Table~\ref{tab:devices}. The schematics are not to scale.}
\label{fig:nozzles}
\end{figure}

A schematic of the experimental setup is shown in Fig.~\ref{fig:schematic}(a).
As a transparent soft-tissue simulant with a shear modulus comparable to that reported for human epidermis, a 5~wt\% gelatin gel (porcine skin gelatin, Sigma-Aldrich) was used~\cite{menezes2009,tagawa2012highly,miyazaki2021dynamic}.
The shear modulus of the gelatin was taken as $G = 5542$~Pa, based on the value reported for a 5~wt\% gelatin gel in a previous study~\cite{kiyama2019gel}.
To prepare the gelatin samples, gelatin powder was dissolved in distilled water to obtain a 5~wt\% solution.
The solution was heated and stirred at 70$^\circ$C and 600~rpm for 30~min until the gelatin was fully dissolved.
The solution was then poured into an acrylic container and stored at 4$^\circ$C for 24~h to allow gelation before the experiments.

The resulting gelatin block (44$\times$44$\times$47~mm$^3$) was placed with its top surface exposed to interact with the jet.
The coordinate system is defined in Fig.~\ref{fig:schematic}(a), where $x$ is the horizontal direction in the image plane, $z$ is the depth direction measured from the gelatin surface, and $y$ is the optical axis of the camera.
A polarization camera (CRYSTA PI-1P, Photron) equipped with a 524~nm band-pass filter was operated at 60{,}000~fps, and images were recorded over a region of 224$\times$512 pixels.
An LED light source (SOLIC-525C, Thorlabs; nominal wavelength $\lambda = 525$~nm) was positioned on the opposite side of the gelatin block.
Figure~\ref{fig:schematic}(b) shows representative raw images and the corresponding binarized cavity images for the two injectors.
The white dashed rectangles indicate the $180\times332$ pixel ROI used for the subsequent stress and cavity-geometry analyses. The cavity depth $h$ was defined as the distance from the gelatin surface to the deepest point of the binarized penetration cavity, whereas the cavity width $w$ was defined as the depth-averaged horizontal extent of the binarized cavity within the ROI. For the Biojector 2000, the V-shaped deformation near the free surface was excluded from these evaluations because it was regarded as surface deformation near the nozzle-contact region rather than part of the main penetration cavity.
Figure~\ref{fig:schematic}(c) illustrates the image-processing procedure used to obtain the binarized cavity image, orientation field, and phase-difference field from the recorded images.
The acquired images were processed and analyzed to obtain the principal stress direction and the corresponding stress intensity.

To generate high-speed liquid jets, two needle-free injectors with different actuation mechanisms and discharge characteristics were used: the Actranza Lab, a pyro-drive jet injector powered by cartridge-based explosive combustion~\cite{miyazaki2019,sonoda2023promising}, and the Biojector 2000, a CO$_2$-driven needle-free injector~\cite{mohammed2010fractional,resik2015immune,sloat2012needle,breitsamer2017needle}.
Miyazaki et al.~\cite{miyazaki2019} reviewed the Actranza Lab, which employs two types of explosives with different burn rates, and reported highly reproducible penetration depth together with high-volume delivery under controllable pressure conditions.
The Biojector 2000 uses pressurized gas to expel the liquid jet at high speed for skin penetration~\cite{bioject_fda2012}.
Although the two devices employ different liquid jet generation mechanisms, this difference enables analysis of stress distributions induced by liquid jets with different discharge characteristics.
The present study focuses not on the superiority of the jet-generation mechanisms themselves, but on the resulting penetration behavior, cavity formation, and stress development in the target material.

The device specifications, including nozzle characteristics, approximate delivery duration, average jet velocity, and average flow rate, are summarized in Table~\ref{tab:devices}, and schematic illustrations of the nozzles are shown in Fig.~\ref{fig:nozzles}.
The delivery duration, $t_d$, was estimated experimentally from the high-speed image sequences as the approximate time interval over which liquid discharge from each injector was observed.
The discharge duration was approximately 10~ms for the Actranza Lab and approximately 27~ms for the Biojector 2000.
Therefore, the values of average flow rate and average jet velocity should be regarded as approximate time-averaged estimates rather than exact instantaneous quantities.
The average flow rate was calculated as
\begin{equation}
Q = \frac{V}{t_d},
\label{eq:flow_rate}
\end{equation}
where $V$ is the injection volume.
Using $V=20~\mu$L, the approximate delivery durations give $Q \approx 2.0$~mL/s for the Actranza Lab and $Q \approx 0.74$~mL/s for the Biojector 2000.
The average jet velocity at the nozzle exit, $U_{\mathrm{avg}}$, was then estimated as
\begin{equation}
U_{\mathrm{avg}} = \frac{Q}{A_n},
\label{eq:jet_velocity}
\end{equation}
where $A_n=\pi d_n^2/4$ is the nozzle cross-sectional area.
The estimated values of $U_{\mathrm{avg}}$ were approximately 250~m/s for the Actranza Lab and approximately 33~m/s for the Biojector 2000.
These values represent approximate time-averaged estimates over the observed discharge duration and do not necessarily correspond to the instantaneous peak jet velocity.

The devices were mounted in a downward orientation, and the nozzle tip was positioned so that it just contacted the upper surface of the gelatin, as shown in the inset of Fig.~\ref{fig:schematic}(a).
This initial condition was selected to ensure firm contact between the nozzle and the gelatin surface while minimizing preload on the target material.
In all experiments, 20~$\mu$L of pure water was injected.
Although the injection volume was fixed, the discharge duration, average jet velocity, and corresponding average flow rate differed between the devices, which, as discussed later, have important implications for the resulting penetration dynamics and stress development.
To assess reproducibility, each condition was repeated at least twice using fresh or undamaged regions of the gelatin block, and consistent trends were obtained.
Although the interaction between the high-speed jet and the gelatin target was complex and destructive, the repeated trials showed qualitatively reproducible penetration behavior and photoelastic stress-response development.

In the following, the principle of the photoelastic measurement used in this study is briefly described.

\subsection{Photoelastic measurement technique}

\begin{figure}[t]
\centering
\includegraphics[width=1.0\textwidth]{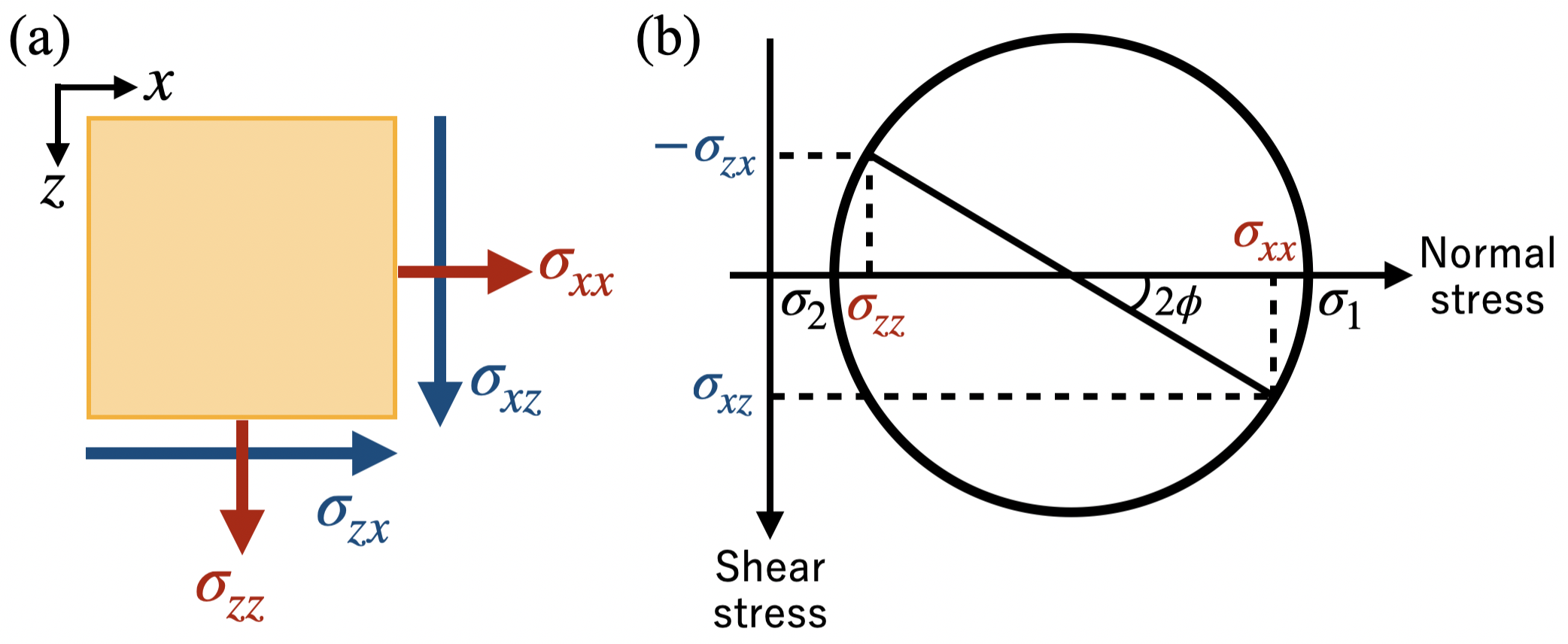}
\caption{(a) Schematic illustration of the normal and shear stresses acting in a two-dimensional plane. (b) Mohr's circle representation of the corresponding two-dimensional stress state.}
\label{fig:Mohr_schematic}
\end{figure}

A high-speed polarization camera operating at 60{,}000~fps was employed to visualize the dynamic stress field within the gelatin during fluid-jet penetration. In the optical setup shown in Fig.~\ref{fig:schematic}(a), the gelatin was placed between the camera and a circularly polarized backlight. When stress is induced in the gelatin, birefringence generates an optical phase difference, $\Delta$, due to the difference in effective refractive indices along the two principal directions~\cite{bass1995handbook}. This phase difference is linearly proportional to the difference in principal stresses, $\sigma_{{d}}$~\cite{yokoyama2026scaling,yokoyama2024high,rastogi2015digital,kulkarni2016optical}:
\begin{equation}
\Delta = C t_p \sigma_{{d}},
\label{eq:Delta1}
\end{equation}
where $C$ is the stress-optical coefficient and $t_p$ is the thickness of the photoelastic material along the optical axis. For the 5~wt\% gelatin used here, $C = 3.825 \times 10^{-8}$~Pa$^{-1}$~\cite{yokoyama2026scaling}, and $t_p = 0.047$~m. For a nonuniform stress field, the measured phase difference represents the optical-path-integrated stress response along the camera axis.

The polarization camera is equipped with a pixelated polarizer array consisting of four adjacent polarizers oriented at $0^\circ$, $45^\circ$, $90^\circ$, and $135^\circ$, enabling simultaneous acquisition of four light intensities ($I_1$, $I_2$, $I_3$, and $I_4$)~\cite{yokoyama2026scaling,yokoyama2024high,onuma2014development}. From these intensities, the instantaneous phase difference $\Delta$ and principal azimuthal angle $\phi$ are obtained as
\begin{equation}
\Delta = \Big( \frac{\lambda}{2 \pi} \Big) \sin^{-1} \frac{\sqrt{(I_3-I_1)^2 + (I_2-I_4)^2}}{I/2},
\label{eq:Delta2}
\end{equation}
\begin{equation}
\phi = \frac{1}{2} \tan^{-1} \frac{I_3-I_1}{I_2-I_4},
\label{eq:phi}
\end{equation}
where $I = I_1+I_2+I_3+I_4$.

To relate the optical quantities to the mechanical stress state around the cavity, the measured phase difference was decomposed into components corresponding to the normal-stress difference and shear stress. Figure~\ref{fig:Mohr_schematic} summarizes the corresponding two-dimensional stress system and its Mohr's circle representation. Based on this stress decomposition and the stress-optic relation, the measured components are expressed as
\begin{equation}
\Delta \cos 2\phi = C \int_{-\infty}^{\infty} (\sigma_{xx}-\sigma_{zz})\,dy,
\label{eq:normal_component}
\end{equation}
\begin{equation}
\Delta \sin 2\phi = 2C \int_{-\infty}^{\infty} \sigma_{xz}\,dy,
\label{eq:shear_component}
\end{equation}
where $y$ denotes the optical axis of the camera, $\sigma_{xx}-\sigma_{zz}$ is the normal-stress difference, and $\sigma_{xz}$ is the shear-stress component. 
Because the photoelastic signal is integrated along the optical path, these quantities should be interpreted as optically integrated, projected components rather than local pointwise stresses. 
In this study, we focus on the magnitudes of the projected stress components. 
Therefore, $\Delta |\cos 2\phi|$ and $\Delta|\sin 2\phi|/2$ are used to represent the projected normal-stress-difference component and the projected shear-stress component, respectively. 
In the following sections, these quantities are referred to as the normal-stress-difference component and the shear-stress component, respectively.


\section{Results and Discussion}
\label{sec:results_discussion}

\subsection{Cavity dynamics}
\begin{figure}[H]
\centering
\includegraphics[width=1.0\linewidth]{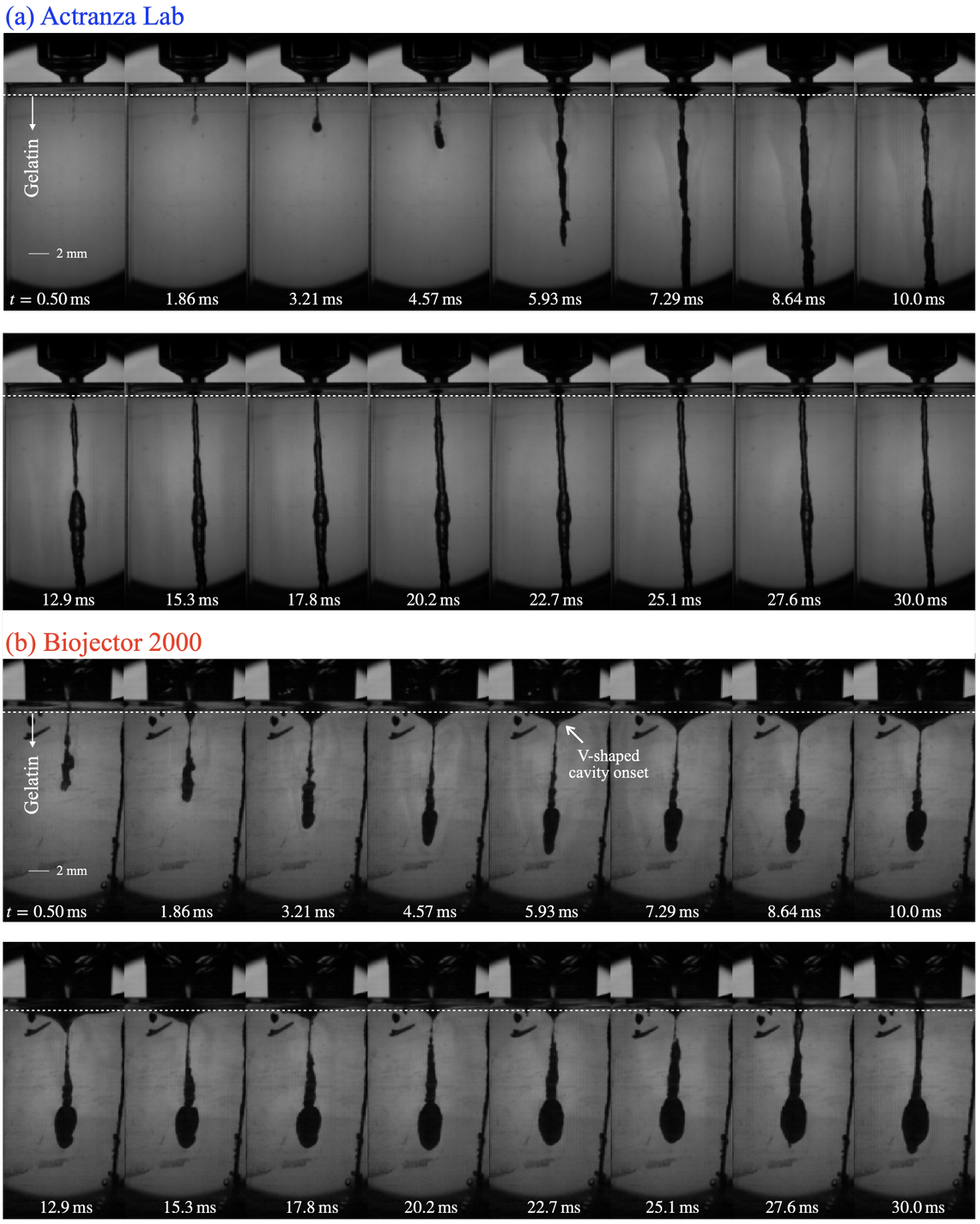}
\caption{
Representative time-sequence images of cavity evolution during liquid-jet penetration into gelatin for (a) the Actranza Lab and (b) the Biojector 2000.
Images are shown from $t = 0.50$ to $30.0~\mathrm{ms}$ after jet impact, with two time windows: $0.50$--$10.0~\mathrm{ms}$ and $12.9$--$30.0~\mathrm{ms}$.
The white dashed line indicates the initial gelatin surface, and the scale bar represents $2~\mathrm{mm}$.
The Actranza Lab forms a relatively slender and elongated cavity, whereas the Biojector 2000 produces a wider cavity accompanied by a characteristic V-shaped deformation near the free surface.
}
\label{fig:Cavity_phenomina}
\end{figure}

Before discussing the photoelastic response, the injector-dependent penetration behavior was first examined from the high-speed images.
Figure~\ref{fig:Cavity_phenomina} shows representative time-sequence images of cavity evolution for the Actranza Lab and the Biojector 2000.
The two devices produced markedly different cavity-development processes.
Overall, the Actranza Lab generated a slender cavity that extended preferentially in the depth direction, whereas the Biojector 2000 produced a wider cavity with a pronounced bulged structure and a characteristic V-shaped deformation near the gelatin surface.

\label{subsec:cavity_phenomena}

For the Actranza Lab, a narrow cavity was observed at the earliest observed time after jet impact and elongated mainly in the vertical direction.
The cavity reached the lower region of the observation area by approximately $5.93$--$7.29~\mathrm{ms}$ while maintaining a relatively small lateral width.
After this stage, the cavity tip was truncated near the lower boundary of the region of interest (ROI), and the cavity depth observed within the image changed only slightly.
The elongated cavity then persisted as a narrow column-like structure throughout the remaining observation period.
This behavior indicates that the injected liquid advanced primarily in the depth direction rather than expanding laterally. The Biojector 2000 exhibited a different penetration mode.
A larger cavity was formed from the early stage of penetration, and the cavity width increased progressively with time.
In addition, a V-shaped deformation near the gelatin surface became visible at approximately $7.29~\mathrm{ms}$.
This V-shaped deformation was regarded as a near-surface deformation around the nozzle-contact region rather than part of the main penetration cavity formed by the liquid jet.
Therefore, it was excluded from the cavity-width and cavity-depth evaluations, and only the main penetration cavity was analyzed.
Compared with the Actranza Lab, the Biojector 2000 promoted lateral cavity expansion and generated a wider, bulged cavity inside the gelatin.

\begin{figure}[t]
\centering
\includegraphics[width=\textwidth]{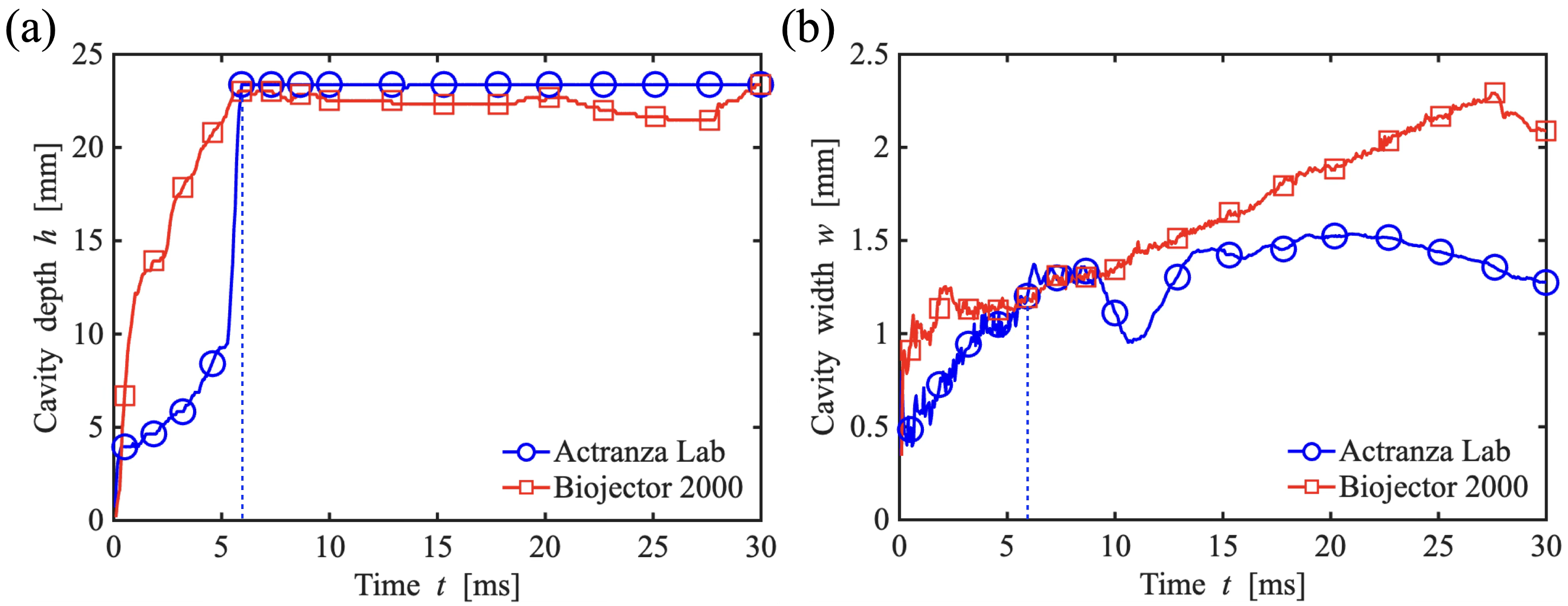}
\caption{
Temporal evolution of the cavity geometry for the Actranza Lab and Biojector 2000.
The plotted data points correspond to the cavity geometries measured from the time-sequence images shown in Fig.~\ref{fig:Cavity_phenomina}.
(a) Time series of the cavity depth $h$, defined as the distance from the gelatin surface to the deepest point of the penetration cavity.
(b) Time series of the cavity width $w$, defined as the average horizontal extent of the binarized penetration cavity computed over the full cavity depth.
In the Biojector 2000 case, the V-shaped deformation near the free surface was excluded from both evaluations because it was regarded as surface deformation near the nozzle-contact region rather than part of the main penetration cavity.
For the Actranza Lab, after the timing indicated by the blue dotted line, the cavity tip had reached the lower boundary of the ROI; therefore, the measured depth represents the maximum depth observable within the ROI rather than the true maximum penetration depth.
}
\label{fig:appendix_wh}
\end{figure}

These qualitative observations were quantified in Fig.~\ref{fig:appendix_wh}.
Figure~\ref{fig:appendix_wh}(a) shows the temporal evolution of the cavity depth $h$.
For the Biojector 2000, the cavity depth increased rapidly immediately after the onset of penetration and reached approximately $22$--$23~\mathrm{mm}$ by around $5~\mathrm{ms}$.
Thereafter, the cavity depth remained nearly constant throughout the observation period.
In contrast, the Actranza Lab showed a more gradual increase in cavity depth during the early stage and reached the maximum observable depth within the ROI at approximately $5.93$--$7.29~\mathrm{ms}$.
It should be noted that, after the timing indicated by the blue dotted line, the cavity tip had reached the lower boundary of the ROI.
Therefore, the measured depth for the Actranza Lab after this point does not represent the true maximum penetration depth, but rather the maximum depth observable within the ROI.

Figure~\ref{fig:appendix_wh}(b) shows the temporal evolution of the cavity width $w$.
For the Actranza Lab, the cavity width increased during the early stage and then remained around $1.3$--$1.5~\mathrm{mm}$, with a slight decrease in the later stage.
In contrast, the Biojector 2000 showed a progressive increase in cavity width within the observation period, eventually exceeding $2~\mathrm{mm}$.
These results indicate that the two injectors exhibited different temporal patterns of lateral cavity expansion.
Although the cavity widths were similar during the early stage, the Actranza Lab maintained a relatively narrow cavity after reaching the lower boundary of the ROI, whereas the Biojector 2000 exhibited greater lateral expansion within the measured time range.
Specifically, the Actranza Lab formed a relatively narrow and depth-oriented cavity, whereas the Biojector 2000 produced a wider cavity and induced greater lateral deformation of the surrounding gelatin during the observed later stage.

Such injector-dependent cavity formation gives rise to distinct mechanical responses inside the gelatin; therefore, the corresponding stress fields are examined next using photoelastic measurements.

\subsection{Stress intensity field}
\begin{figure}[H]
\centering
\includegraphics[width=\textwidth]{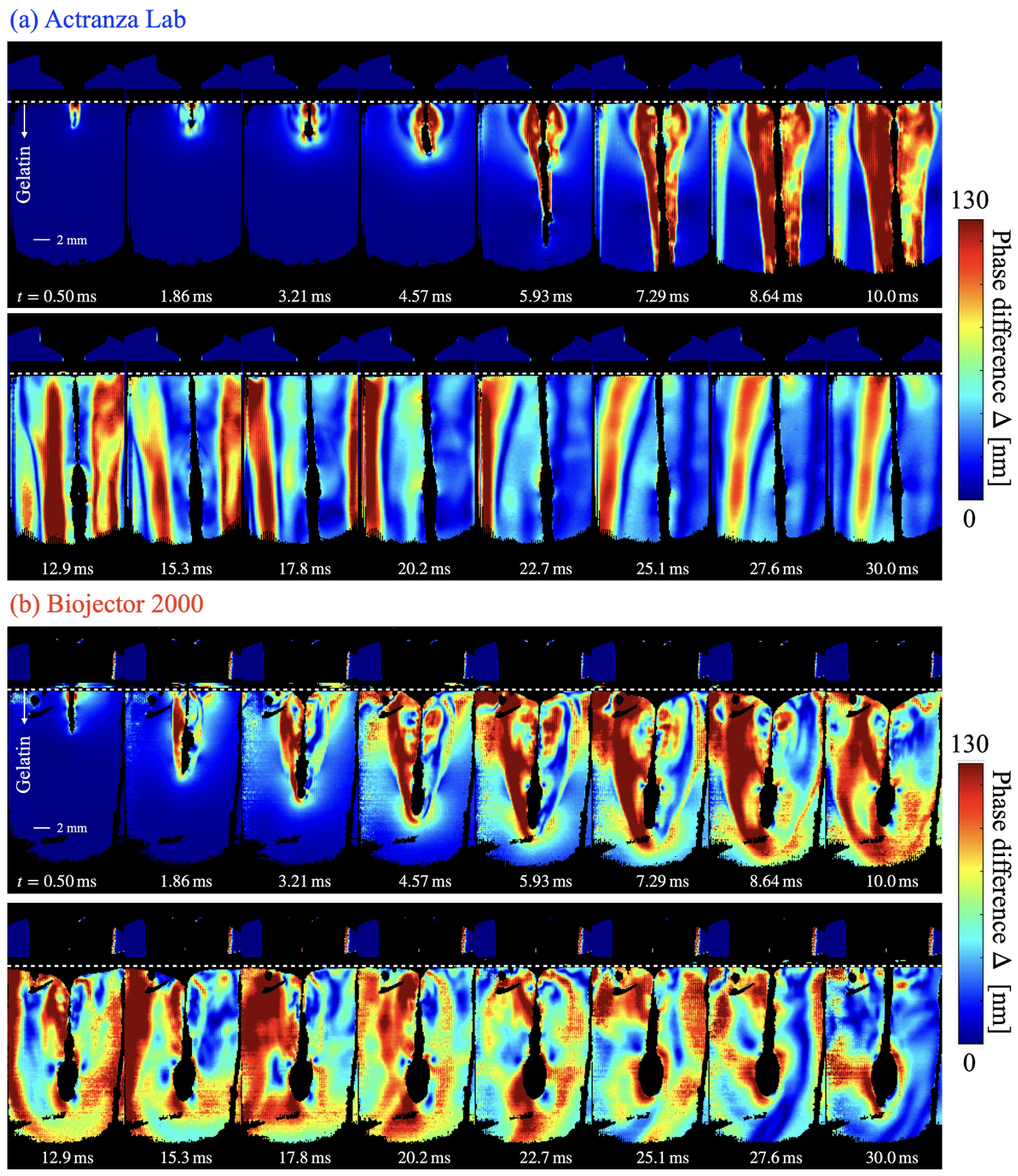}
\caption{
Temporal evolution of the photoelastic stress-intensity responses surrounding the penetration trace induced by liquid-jet injection using (a) the Actranza Lab and (b) the Biojector 2000.
Images are shown from $t = 0.50$ to $30.0~\mathrm{ms}$ after jet impact, with two time windows: $0.50$--$10.0~\mathrm{ms}$ and $12.9$--$30.0~\mathrm{ms}$.
The total injected volume was 20~$\mu$L for both devices.
The regions corresponding to the liquid-jet penetration trace were masked in black, and the phase-difference field in the surrounding gelatin was used to evaluate the photoelastic stress-intensity response distribution.
The white dashed line indicates the initial gelatin surface, and the scale bar represents $2~\mathrm{mm}$.
The phase difference is displayed using the same color scale for both devices over the range of 0--130~nm.
}
\label{fig:stressfields}
\end{figure}

The temporal evolution of the photoelastic stress response in gelatin was analyzed and compared for the two needle-free injection devices, the Actranza Lab and the Biojector 2000. As described in Sect.~\ref{sec:expt}, 20~$\mu$L of pure water was injected into the gelatin with the nozzle tip in contact with the gelatin surface. Figure~\ref{fig:stressfields} shows the resulting time evolution of the photoelastic stress-intensity response induced by liquid-jet penetration. In these images, the penetration trace formed by the liquid jet was masked in black, and the photoelastic response generated around this region was evaluated. The phase-difference field shown by the color map represents the spatial distribution of the photoelastic stress-intensity response. According to Eq.~\eqref{eq:Delta1}, a larger phase difference corresponds to a larger principal stress difference.

Figure~\ref{fig:stressfields} reveals that the spatial distribution and temporal evolution of the photoelastic stress-response field differed markedly between the Actranza Lab and the Biojector 2000. For the Actranza Lab, high photoelastic stress-intensity response initially developed around the narrow and deep penetration trace and subsequently propagated into the surrounding region. In the early stage, the high-phase-difference region was mainly concentrated near the penetration trace and close to the gelatin surface. In contrast, the Biojector 2000 generated a broader photoelastic stress response from the relatively early stage of penetration. As the bulged penetration trace developed, the high-phase-difference region expanded over a wider portion of the observation area. Thus, the Actranza Lab produced a localized photoelastic stress response concentrated along the deep penetration direction, whereas the Biojector 2000 produced a more broadly distributed photoelastic stress response.
\label{subsec:stress_intensity}

\begin{figure}[t]
\centering
\includegraphics[width=\textwidth]{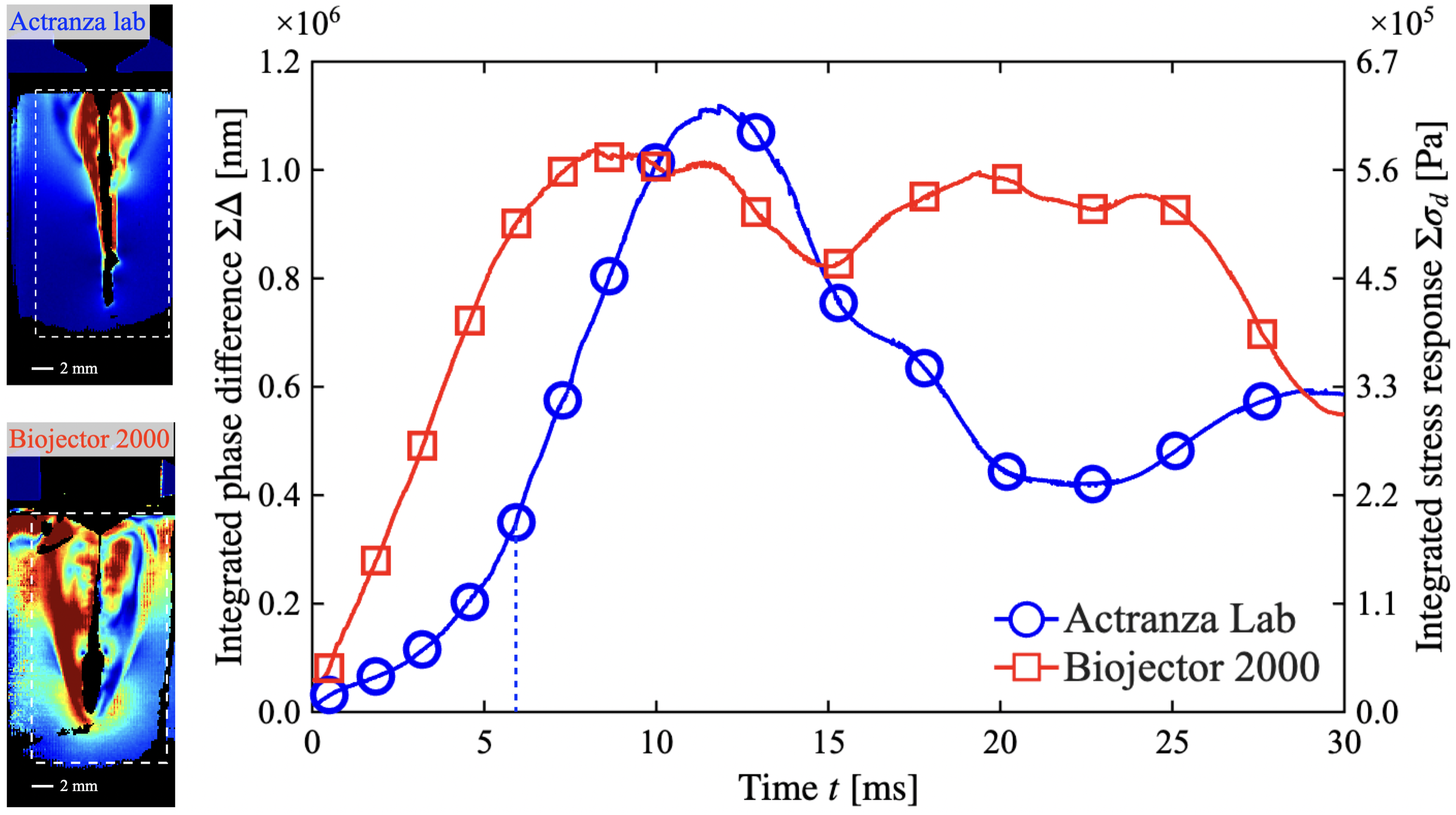}
\caption{
Temporal evolution of the spatially integrated phase difference for the Actranza Lab and Biojector 2000, each using a total injected volume of 20~$\mu$L.
The left panels show the photoelastic response images with the 180$\times$332 pixel region of interest (ROI) used for spatial integration, indicated by the white dashed rectangles.
The phase difference was spatially integrated over this ROI; therefore, the plotted value represents the total photoelastic response within the ROI rather than a local stress magnitude or maximum phase difference.
The markers on the temporal profiles correspond to the time points shown in Fig.~\ref{fig:stressfields}.
For the Actranza Lab, the blue dotted line indicates the timing after which the cavity tip had reached the lower boundary of the ROI, corresponding to the termination of observable penetration within the ROI.
According to Eq.~\eqref{eq:Delta1}, the optical phase difference $\Delta$ is linearly proportional to the principal stress difference $\sigma_d$.
For reference, the corresponding integrated stress-response scale on the right-hand axis was estimated using $C = 3.825 \times 10^{-8}$~Pa$^{-1}$~\cite{yokoyama2026scaling} and $t_p = 0.047~\mathrm{m}$ for the 5~wt\% gelatin used in this study.
This right-hand scale is provided only as a relative stress-response scale and does not represent a local stress magnitude.
}
\label{fig:comparison}
\end{figure}

For quantitative comparison of these spatial differences, a rectangular region of interest (ROI), indicated by the white dashed rectangles in the left panels of Fig.~\ref{fig:comparison}, was defined, and the phase difference was spatially integrated within this region. The ROI size was fixed at 180$\times$332 pixels for both devices so that the temporal evolution of the photoelastic response could be compared over the same observation area. Therefore, the integrated phase difference shown in Fig.~\ref{fig:comparison} represents the total photoelastic response within the ROI rather than a local stress magnitude or maximum phase difference. For the Actranza Lab, Fig.~\ref{fig:comparison} shows that the integrated phase difference increased gradually after jet impact, continued to increase even after the termination of observable penetration, and reached a maximum at approximately $t \approx 12$--13~ms before decreasing. This behavior indicates that, for the Actranza Lab, the local photoelastic response first developed around the narrow and deep penetration trace, whereas the integrated response within the ROI became largest only after the photoelastic response field had spread over a broader area. In contrast, the integrated phase difference for the Biojector 2000 increased rapidly from the early stage, reached a high level at around $t \approx 8$--10~ms, and then remained high for an extended period. This trend reflects the formation and persistence of a broad stressed region associated with the wide penetration trace and bulged structure observed in Fig.~\ref{fig:stressfields}.

Overall, although the maximum integrated phase difference was similar for the two devices, the temporal evolution and spatial distribution of the photoelastic stress response differed markedly. The Actranza Lab generated a more localized stress response associated with a narrow, deep penetration trace and a delayed peak in the integrated response, whereas the Biojector 2000 produced a broader stress distribution and a sustained high integrated response owing to the formation of a wide penetration trace and bulged structure. These results indicate that injector-dependent penetration dynamics strongly affect the spatial and temporal development of the photoelastic stress-response field.

In the following sections, the photoelastic response obtained here is further decomposed into stress components, and the projected shear-stress and normal-stress-difference components are examined separately to clarify the relative contributions of different stress modes. This analysis enables a more detailed evaluation of which stress components are most strongly affected by injector-dependent penetration dynamics.

\subsection{Shear stress during penetration}
\label{subsec:shear_stress}
High-speed liquid-jet penetration into soft materials has often been modeled or interpreted in relation to deformation around the penetration cavity~\cite{shergold2006,tagawa2013needle,miyazaki2021dynamic,mousavi2025modelling,rossello2022bullet,yamagata2026penetration}. 
Because the deformation around an elongated penetration cavity can include substantial shear components, the projected shear-stress component was evaluated in the present study.
The projected shear-stress component was evaluated on the basis of Eq.~\ref{eq:shear_component} by separating it from the experimentally measured phase difference $\Delta$ and principal stress direction $\phi$.

\begin{figure}[H]
\centering
\includegraphics[width=\textwidth]{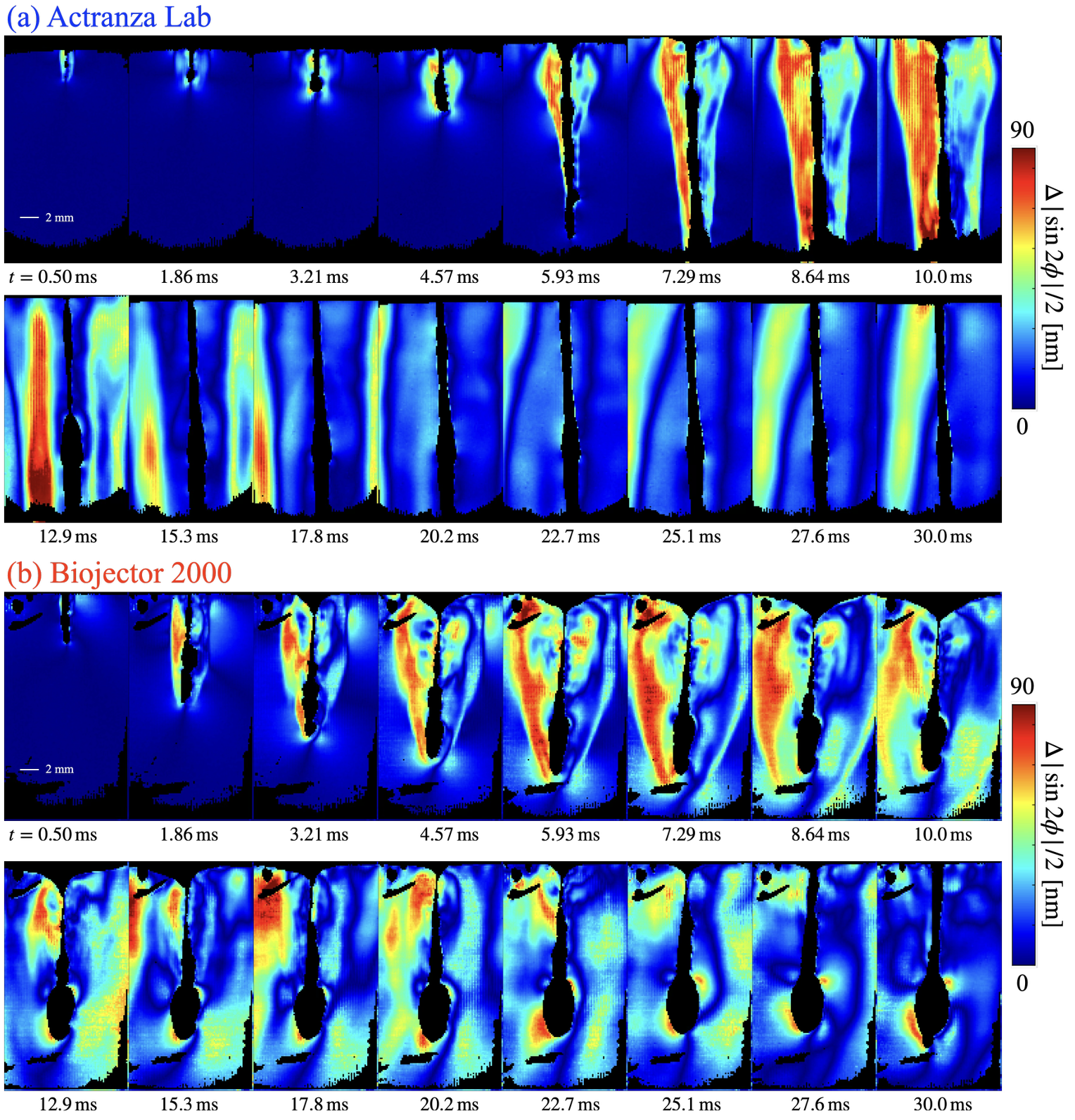}
\caption{
Temporal evolution of the spatial distribution of the projected shear-stress component within the region of interest (ROI) during high-speed liquid-jet penetration for (a) the Actranza Lab and (b) the Biojector 2000.
Images are shown from $t = 0.50$ to $30.0~\mathrm{ms}$ after jet impact, with two time windows: $0.50$--$10.0~\mathrm{ms}$ and $12.9$--$30.0~\mathrm{ms}$.
The projected shear-stress component was calculated from the experimentally measured phase difference $\Delta$ and principal stress direction $\phi$, as described in Eq.~\eqref{eq:shear_component}.
Because the present analysis focuses on the magnitude of the projected shear-stress component, the color scale represents $\Delta|\sin 2\phi|/2$ and is displayed over the same range of 0--90 nm for both devices.
The regions corresponding to the liquid-jet penetration trace were masked in black.
}
\label{fig:Shear_stress_colormap}
\end{figure}

Figure~\ref{fig:Shear_stress_colormap} shows the temporal evolution of the spatial distribution of the projected shear-stress component visualized by photoelastic measurement. 
The quantity shown here, $\Delta |\sin 2\phi|/2$, represents the photoelastic signal corresponding to the projected shear-stress component, obtained from the measured phase difference $\Delta$ and principal stress direction $\phi$.

In the Actranza Lab case, the projected shear-stress component was initially observed near the jet-impact region and the upper part of the penetration path. 
As penetration proceeded, the response appeared along the cavity region and gradually extended over a larger portion of the ROI. 
The response was not spatially uniform, and the location and extent of the local high-response region changed with time. In the Biojector 2000 case, the projected shear-stress component also appeared around the penetration region from the early stage. 
The response region subsequently changed with time and exhibited a spatial distribution different from that observed in the Actranza Lab case. 
This indicates that the shear-related photoelastic response is affected not only by the penetration depth but also by the cavity formation and transient deformation process generated by each injector. 
Thus, Fig.~\ref{fig:Shear_stress_colormap} shows that the projected shear-stress component develops both temporally and spatially within the material during jet penetration.

\begin{figure}[t]
\centering
\includegraphics[width=\textwidth]{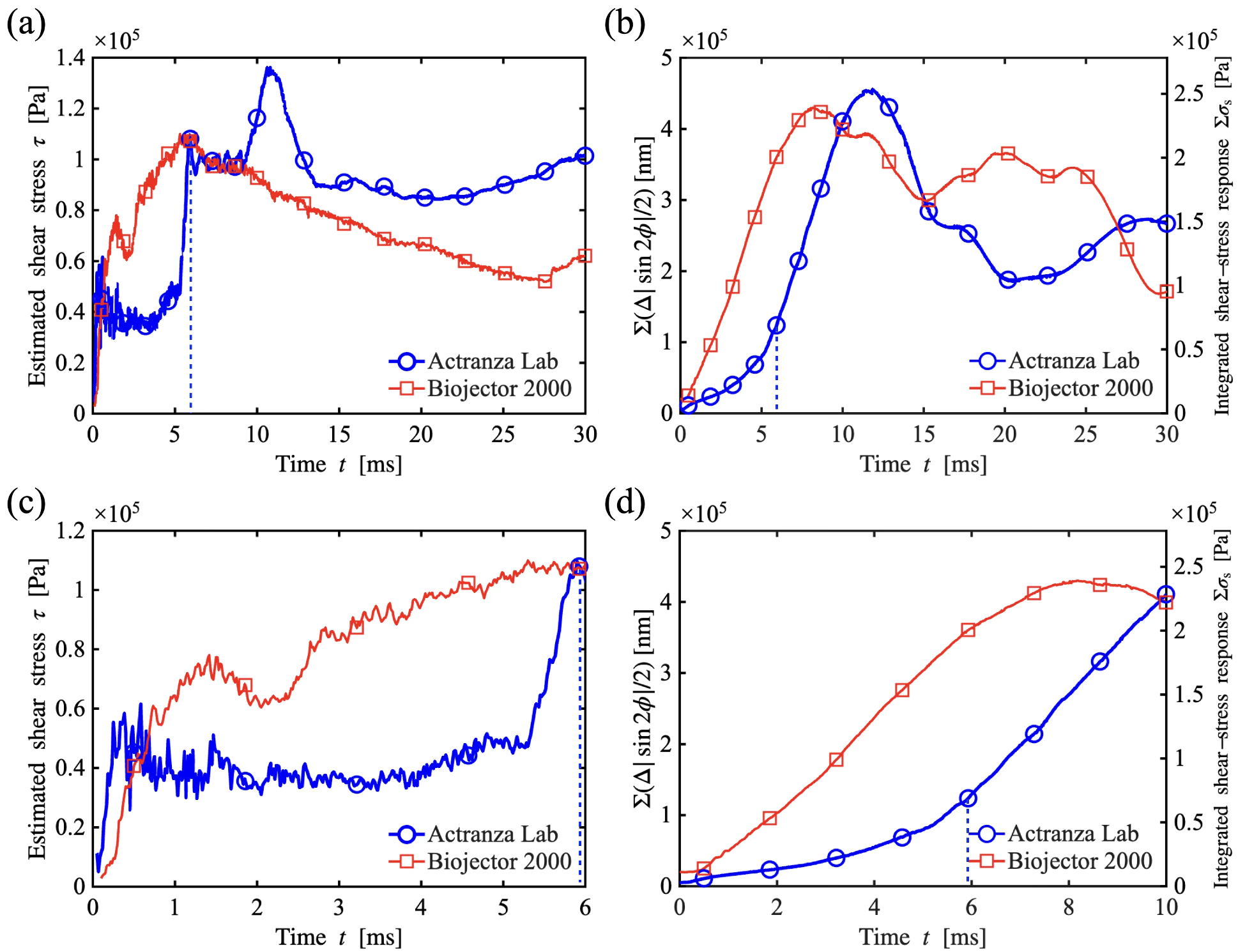}
\caption{
Comparison of the temporal evolution of the shear response during liquid-jet penetration for the Actranza Lab and Biojector 2000.
(a) Estimated representative shear stress $\tau$ obtained from the cavity depth $h$ and average cavity width $w$.
(b) Spatially integrated magnitude of the projected shear-response component, $\Sigma(\Delta|\sin 2\phi|/2)$, over the ROI; the corresponding integrated shear-stress-response scale is shown on the right-hand axis.
(c) Early-stage view of the estimated representative shear stress $\tau$ up to $6~\mathrm{ms}$.
(d) Early-stage view of the spatially integrated magnitude of the projected shear-response component, $\Sigma(\Delta|\sin 2\phi|/2)$, up to $10~\mathrm{ms}$; the corresponding integrated shear-stress-response scale is shown on the right-hand axis.
The plotted markers in (b) and (d) correspond to the time-sequence images shown in Fig.~\ref{fig:Shear_stress_colormap}.
The blue dotted lines indicate the timing after which the Actranza Lab cavity tip had reached the lower boundary of the ROI, corresponding to the termination of observable penetration within the ROI.
The right-hand scale is provided for relative comparison and should not be interpreted as a local shear-stress magnitude.
}
\label{fig:Graph_stress}
\end{figure}

Previous studies have investigated the penetration of high-speed liquid jets into soft materials and skin-simulating gels~\cite{rossello2022bullet,yamagata2026penetration}.
Here, the temporal evolution of the shear response was compared using two approaches: a cavity-geometry-based estimate and the photoelastic measurement performed in the present study. 
Figure~\ref{fig:Graph_stress}(a) shows the representative shear stress estimated from the cavity geometry, following the shear-deformation-based interpretation proposed for liquid-jet penetration into gelatin~\cite{yamagata2026penetration}.
The cavity depth $h$ and average cavity width $w$ were evaluated from binarized high-speed images, as detailed in Sect.~\ref{subsec:cavity_phenomena}. 
The representative shear stress was estimated as
\begin{equation}
\tau = G \frac{h}{w},
\label{eq:estimated_shear_stress}
\end{equation}
where $h$ denotes the cavity depth and $w$ denotes the average cavity width. 
The shear modulus of gelatin was set to $G = 5542\,\mathrm{Pa}$, using the material property reported in Ref.~\cite{kiyama2019gel}.

Figure~\ref{fig:Graph_stress}(b) shows the spatially integrated shear-stress response obtained by photoelastic measurement. 
This value was calculated by integrating the projected shear-stress component corresponding to $\Delta|\sin 2\phi|/2$ over the ROI. 
Thus, Fig.~\ref{fig:Graph_stress}(a) represents a geometry-based estimate of the representative shear stress, whereas Fig.~\ref{fig:Graph_stress}(b) represents the total shear-related photoelastic response generated within the ROI.

First, comparison between Fig.~\ref{fig:Graph_stress}(a) and Fig.~\ref{fig:Graph_stress}(b) shows that, although the two quantities are not identical, their overall temporal trends show a certain correspondence. 
In particular, for the Actranza Lab, both the estimated shear stress and the spatially integrated photoelastic response increased after the onset of penetration and reached large values around $13~\mathrm{ms}$. 
This suggests that the cavity-geometry-based estimate captures the overall timescale of the shear-related response observed inside the material. However, this correspondence in the overall trend does not necessarily mean that the two responses evolve in the same manner during the initial stage. 
When comparing the photoelastic shear response with the geometry-based estimate, it is important to use a time range in which the entire penetration trace is captured within the ROI. 
After the penetration trace extends outside the ROI, part of the photoelastic response may exist outside the analysis region, making the comparison based on the spatially integrated value incomplete. 
Therefore, in this study, we focused on the early stage up to $6~\mathrm{ms}$, during which the entire penetration trace was contained within the ROI, as enlarged in Fig.~\ref{fig:Graph_stress}(c) and Fig.~\ref{fig:Graph_stress}(d).

Figure~\ref{fig:Graph_stress}(c) shows the shear stress estimated from the cavity geometry during the early stage. 
In this estimate, the shear stress starts to increase immediately after the onset of penetration, indicating a rapid response to changes in the cavity depth $h$ and average cavity width $w$. 
This is because the geometry-based estimate directly reflects the local cavity deformation generated by the jet. In contrast, the spatially integrated shear-stress response obtained by photoelastic measurement, shown in Fig.~\ref{fig:Graph_stress}(d), exhibits a more gradual initial rise. 
Compared with the geometry-based estimate in Fig.~\ref{fig:Graph_stress}(c), the increase in the photoelastic response does not simply follow the instantaneous change in cavity geometry. 
In other words, although the two responses show a similar overall temporal trend in the full-time comparison, the photoelastic measurement reveals that the internal shear-related response develops more gradually during the early stage. 
This suggests that the local deformation generated near the jet is reflected in the photoelastic stress-response field over a finite timescale as the response field develops and spreads within the ROI.

These results indicate that the cavity-geometry-based estimate and the photoelastic measurement provide complementary information. 
The comparison between Fig.~\ref{fig:Graph_stress}(a) and Fig.~\ref{fig:Graph_stress}(b) shows that the two responses exhibit similar overall temporal behavior, including the increase toward the large response around $13~\mathrm{ms}$. 
On the other hand, the early-stage comparison between Fig.~\ref{fig:Graph_stress}(c) and Fig.~\ref{fig:Graph_stress}(d) reveals a difference in the initial rise: the photoelastic response develops more gradually than the cavity-geometry-based estimate. 
This difference suggests that the measured photoelastic shear-related response develops over a finite timescale, rather than following the cavity deformation instantaneously.
Therefore, although the cavity-geometry-based estimate is useful for evaluating a representative shear load associated with penetration, the temporal development of the photoelastic stress-response field must also be considered when discussing the mechanical stimulus applied to biological tissues or tissue simulants during high-speed jet injection.

\subsection{Comparison of shear and normal-stress-difference responses}
\label{subsec:comparison_shear_normal}

Although the preceding section focused on the projected shear-stress component, the photoelastic stress-response field generated during liquid-jet penetration cannot be described solely by the shear-related component. 
To further characterize the internal mechanical response, we evaluated the projected normal-stress-difference component, represented by $\Delta |\cos 2\phi|$, in addition to the projected shear-stress component, $\Delta|\sin 2\phi|/2$. 
Here, $\Delta |\cos 2\phi|$ should be interpreted as an optically derived projected normal-stress-difference component obtained from the measured phase difference $\Delta$ and principal stress direction $\phi$, rather than as a local pointwise normal stress.

\begin{figure}[H]
\centering
\includegraphics[width=\textwidth]{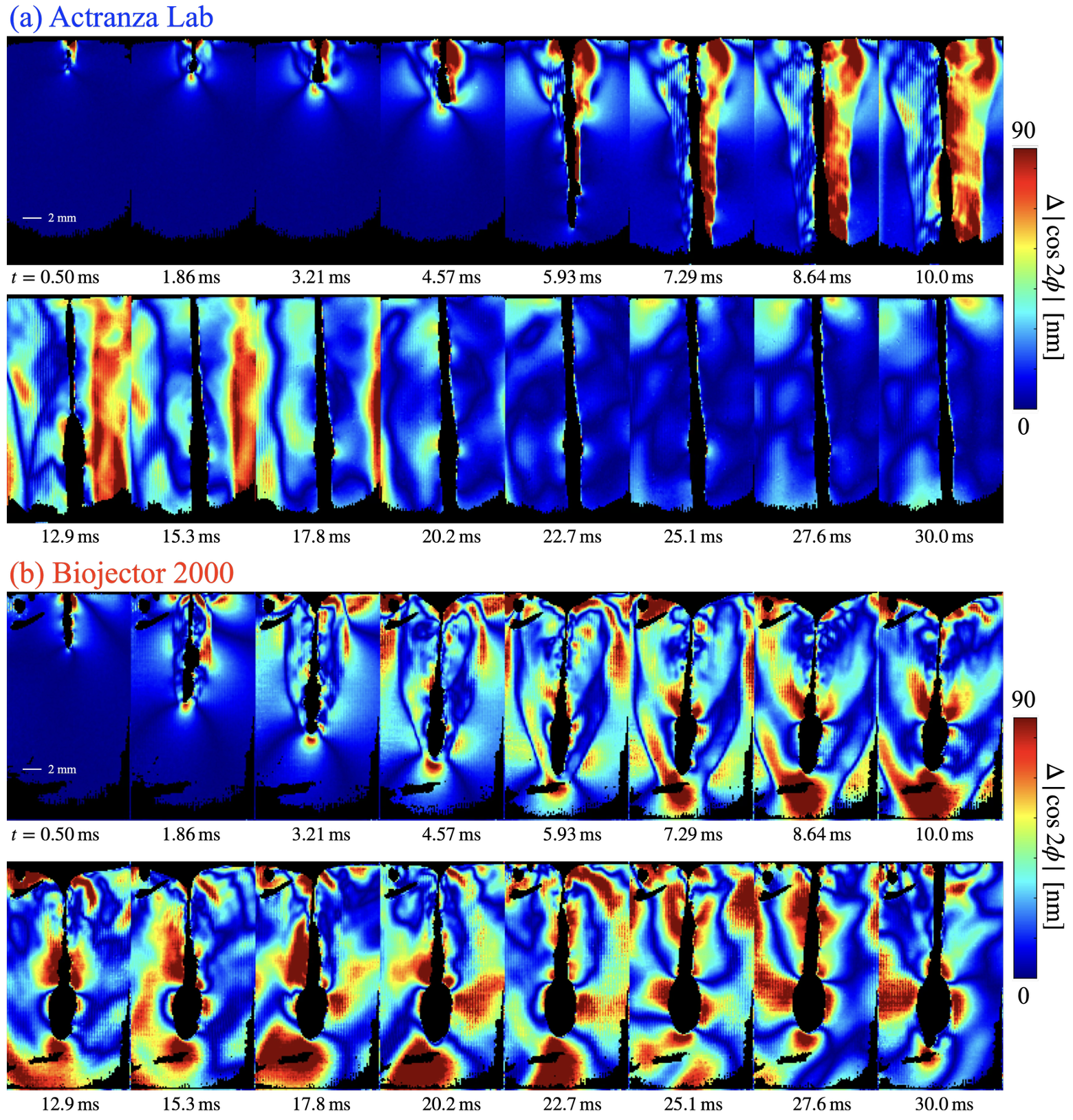}
\caption{ Temporal evolution of the spatial distribution of the projected normal-stress-difference component within the region of interest (ROI) during high-speed liquid-jet penetration for (a) the Actranza Lab and (b) the Biojector 2000. Images are shown from $t = 0.50$ to $30.0~\mathrm{ms}$ after jet impact, with two time windows: $0.50$--$10.0~\mathrm{ms}$ and $12.9$--$30.0~\mathrm{ms}$. The projected normal-stress-difference component was calculated from the experimentally measured phase difference $\Delta$ and principal stress direction $\phi$, as described in Eq.~\eqref{eq:normal_component}. Because the present analysis focuses on the magnitude of the projected normal-stress-difference component, the color scale represents $\Delta|\cos 2\phi|$ and is displayed over the same range of 0--90~nm for both devices. The regions corresponding to the liquid-jet penetration trace were masked in black. }
\label{fig:normal_stress_colormap}
\end{figure}

Figure~\ref{fig:normal_stress_colormap} shows the temporal evolution of the spatial distribution of $\Delta |\cos 2\phi|$ during high-speed jet penetration for the Actranza Lab and Biojector 2000. 
For the Actranza Lab, the response initially appeared near the jet-impact region and along the slender penetration path. 
As penetration proceeded, the projected normal-stress-difference component developed around the elongated cavity while remaining relatively localized along the depth-oriented penetration region. 
In contrast, for the Biojector 2000, the $\Delta |\cos 2\phi|$ distribution expanded over a broader region of the ROI, particularly during the later stage when lateral cavity expansion and bulging were observed. 
This indicates that the cavity morphology produced by each injector is closely related not only to the shear-related response but also to the development of the normal-stress-difference component.

\begin{figure}[t]
\centering
\includegraphics[width=\textwidth]{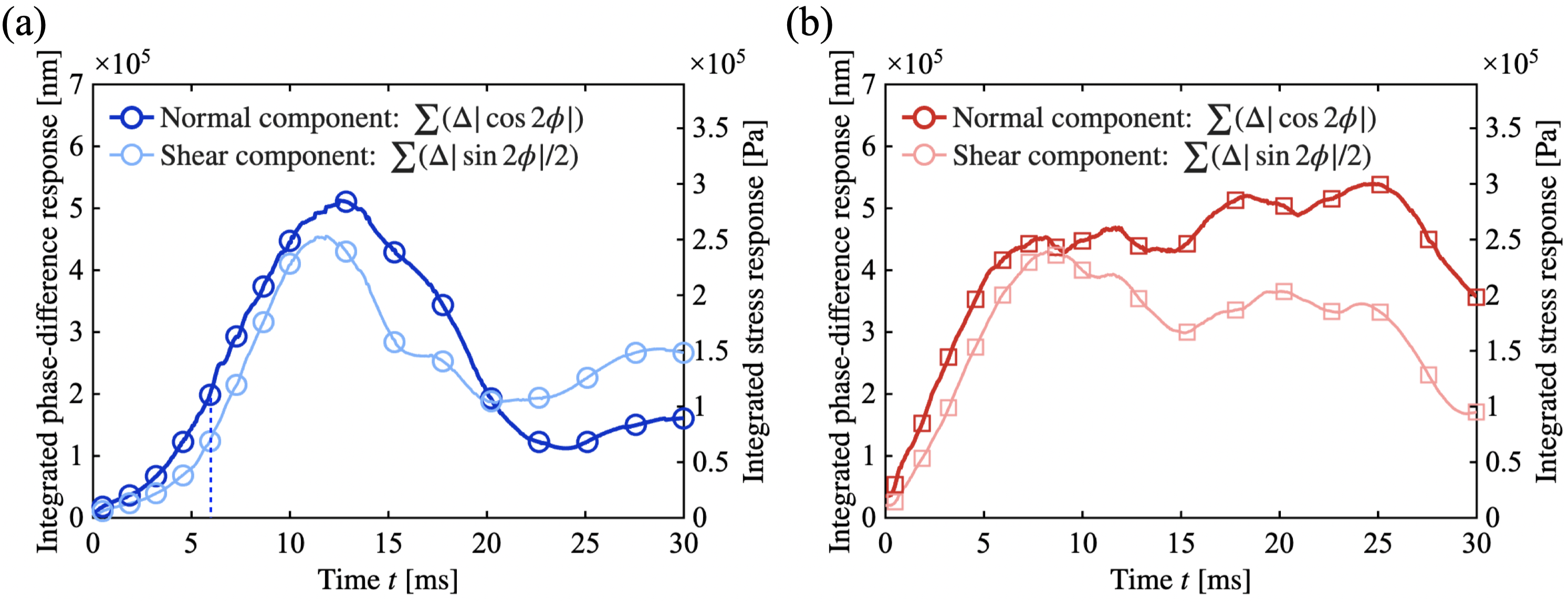}
\caption{
Temporal evolution of the integrated phase-difference components obtained from photoelastic measurements for the Actranza Lab and Biojector 2000.
The magnitude of the projected normal-stress-difference component is represented by $\sum(\Delta|\cos 2\phi|)$, whereas the magnitude of the projected shear-stress component is represented by $\sum\left(\Delta|\sin 2\phi|/2\right)$.
The summation was performed over the ROI.
The left vertical axis represents the integrated phase-difference response [nm].
The right vertical axis represents the corresponding integrated stress-response scale converted using the photoelastic coefficient.
This right-hand scale is provided for relative comparison and should not be interpreted as a local stress magnitude.
}
\label{fig:Normal_Shear_comparison}
\end{figure}

Figure~\ref{fig:Normal_Shear_comparison} compares the temporal evolution of the spatially integrated phase-difference components within the ROI. 
The projected normal-stress-difference response was evaluated as $\sum (\Delta |\cos 2\phi|)$, whereas the projected shear-stress response was evaluated as $\sum (\Delta |\sin 2\phi|/2)$. 
The left vertical axis represents the integrated phase-difference response [nm], and the right vertical axis represents the corresponding integrated stress response [Pa] converted using the photoelastic coefficient. 
Thus, the plotted quantities represent spatially integrated photoelastic responses within the ROI, rather than local stress values.

For the Actranza Lab, both $\sum (\Delta |\cos 2\phi|)$ and $\sum (\Delta |\sin 2\phi|/2)$ increased after the onset of penetration and reached large values while the slender cavity was developing. 
Although the response subsequently decreased, both components remained finite during the later stage. 
For the Biojector 2000, both components increased rapidly in the early stage and remained relatively large over a longer period. 
In particular, the sustained magnitude of $\sum (\Delta |\cos 2\phi|)$ is consistent with the broad distribution of $\Delta |\cos 2\phi|$ observed in Fig.~\ref{fig:normal_stress_colormap}. 
These results show that the normal-stress-difference component is not a negligible secondary contribution, but is present at an order comparable to the shear-stress component.

This finding is important because liquid-jet penetration and needle-free injection have often been discussed mainly in terms of cavity deformation and shear-related mechanical effects~\cite{{shergold2006,tagawa2013needle,miyazaki2021dynamic,mousavi2025modelling,rossello2022bullet,yamagata2026penetration}}. 
In addition, shear stress has been considered an important factor in the biological response and delivery mechanism of pyro-drive jet injection, as discussed by Sonoda et al.~\cite{sonoda2023promising}. 
However, the present photoelastic measurements demonstrate that, during jet penetration, the normal-stress-difference component can develop with a magnitude comparable to that of the shear-stress component. 
Therefore, the internal mechanical response induced by liquid-jet penetration should not be interpreted solely as a shear-stress-dominated response. 
Rather, it should be regarded as a multi-component stress response in which both the shear-stress component and the normal-stress-difference component develop in association with injector-dependent cavity dynamics.

These results suggest that the mechanical evaluation of needle-free injection devices should consider not only shear stress but also the normal-stress-difference component. 
This is particularly important for injectors such as the Biojector 2000, where lateral cavity expansion and bulging generate a broad and sustained internal stress response. 
Considering both stress components provides a more complete framework for understanding tissue deformation and mechanical stimulation induced by high-speed liquid-jet injection.

\section{Conclusion}
\label{sec:conclusion}
In this study, high-speed photoelastic measurements were used to visualize and quantify optically integrated stress responses generated in a 5~wt\% gelatin tissue simulant during needle-free jet penetration by two injectors with different actuation mechanisms. Although both devices injected the same liquid volume, they produced markedly different cavity dynamics. The Actranza Lab formed a relatively narrow and depth-oriented cavity that reached the lower boundary of the observable region, whereas the Biojector 2000 formed a wider, bulged cavity that remained within the observation region. These injector-dependent cavity dynamics resulted in distinct spatial and temporal developments of the photoelastic stress-response field.

The measured phase difference was decomposed into projected shear-stress and normal-stress-difference components. The results showed that the mechanical response during jet penetration cannot be interpreted solely in terms of shear stress. For the Actranza Lab, the normal-stress-difference component developed around the slender cavity and became comparable to the shear-stress component during part of the penetration process. For the Biojector 2000, this component was broadly distributed and remained large during the later stage associated with lateral cavity expansion and bulging. Thus, under the present experimental conditions, the normal-stress-difference component was not a negligible secondary contribution, but an essential part of the stress response generated during liquid-jet penetration.

These findings extend the conventional shear-centered interpretation of needle-free jet penetration and show that cavity formation can generate a multi-component photoelastic stress response whose spatial extent, duration, and dominant component depend on injector-specific cavity dynamics. From a biomedical engineering perspective, this stress-based description provides a quantitative framework for evaluating tissue loading during needle-free injection, beyond penetration depth alone. Although the present study used a gelatin tissue simulant and did not directly quantify biological outcomes, the measured projected stress-response topology may help interpret injector-dependent differences in tissue deformation, pain-related mechanical stimulation, tissue damage, and delivery distribution. Future studies combining this photoelastic stress-response analysis with biological measurements will be useful for linking injector design, mechanical tissue loading, and biomedical performance.

\backmatter

\bmhead{Acknowledgements}
The authors thank the members of the Tagawa laboratory for helpful discussions and technical support.

\section*{Declarations}

\textbf{Funding}\\
This study was supported by collaborative research funding from Daicel Corporation.
\textbf{Competing interests}\\
This study was conducted as a collaborative research project with Daicel Corporation. The Actranza Lab injector evaluated in this study was developed by Daicel Corporation, and Kazuhiro Terai is an employee of Daicel Corporation. The authors declare no other financial or non-financial competing interests related to this study.

\textbf{Ethics approval and consent to participate}\\
Not applicable.

\textbf{Consent for publication}\\
Not applicable.

\textbf{Data availability}\\
The data that support the findings of this study are available from the corresponding author upon reasonable request.

\textbf{Materials availability}\\
Not applicable.

\textbf{Code availability}\\
Not applicable.

\textbf{Author contribution}\\
Kohei Yamagata: Investigation, data analysis, visualization, interpretation, writing---original draft, review, and editing.\\
Prasad Sonar: Investigation, data analysis, interpretation, writing---original draft, review, and editing.\\
Kazuhiro Terai: Investigation, interpretation, writing---review and editing.\\
Yuto Yokoyama: Photoelastic methodology, interpretation, writing---review and editing.\\
Yoshiyuki Tagawa: Conceptualization, supervision, funding acquisition, writing---review and editing.
\begin{appendices}

\section{Derivation of the decomposed photoelastic components}
\label{app:decomposition}

Under plane-stress conditions, the in-plane stress tensor is written as
\begin{equation}
\boldsymbol{\sigma}
=
\begin{pmatrix}
\sigma_{xx} & \sigma_{xz} \\
\sigma_{xz} & \sigma_{zz}
\end{pmatrix}.
\end{equation}
The principal stress difference is expressed as
\begin{equation}
\sigma_d
=
\sqrt{(\sigma_{xx}-\sigma_{zz})^2 + (2\sigma_{xz})^2},
\end{equation}
and the principal direction satisfies
\begin{equation}
\tan 2\phi = \frac{2\sigma_{xz}}{\sigma_{xx}-\sigma_{zz}}.
\end{equation}
Accordingly,
\begin{equation}
\sigma_d \cos 2\phi = \sigma_{xx}-\sigma_{zz},
\end{equation}
and
\begin{equation}
\sigma_d \sin 2\phi = 2\sigma_{xz}.
\end{equation}

Using the stress-optic relation, the measured phase difference is proportional to the line-of-sight integral of the principal stress difference. Therefore, the experimentally obtained quantities $\Delta \cos 2\phi$ and $\Delta \sin 2\phi$ correspond to the normal-stress-difference-related and shear-stress-related components, respectively. The signed quantities $\Delta \cos 2\phi$ and $\Delta \sin 2\phi/2$ correspond to the projected normal-stress-difference-related and shear-stress-related components, respectively. 
In the quantitative analyses in the main text, their magnitudes, $\Delta|\cos 2\phi|$ and $\Delta|\sin 2\phi|/2$, were used to compare the spatial distributions and integrated responses without cancellation between positive and negative contributions.

Because the present measurements are based on optical integration along the camera axis, these quantities should be interpreted as projected stress-related components rather than strictly local pointwise stresses. Nevertheless, they provide a consistent basis for comparing the spatiotemporal development of the stress field between the two injectors.

\end{appendices}

\end{document}